\documentclass[12pt]{iopart}
\usepackage{graphicx}
\usepackage{iopams}
\usepackage{amstext}
\usepackage{bm}
\usepackage{url}

\newcommand{\xtenBfiveOpt}{112}
\newcommand{\xtenBOpt}{116}
\newcommand{\xtenBfiveGap}{0.001}
\newcommand{\xtenBGap}{0.001}

\begin{document}

\title[Lehmer-code VQKAN for the traveling salesman problem]{Solving the traveling salesman problem on $\lceil \log_2 (N-1)! \rceil$ qubits with a variational quantum Kolmogorov--Arnold network: exact-gradient CVaR training and generalization over time-dependent graphs}

\author{Hikaru Wakaura}
\address{QuantScape Inc., 4-11-18, Manshon-Shimizudai, Meguro, Tokyo, 153-0064, Japan}
\ead{hikaruwakaura@gmail.com}

\begin{abstract}
The one-hot encoding of the traveling salesman problem (TSP) on quantum hardware needs $(N-1)^2$ qubits for $N$ sites and penalty terms that dominate the cost landscape.
Permutation encodings that label every tour by its Lehmer code need only $\lceil \log_2 (N-1)! \rceil$ qubits and no penalty term, but variational solvers on them have reached the optimum reliably only up to six sites.
Here we show how to train such a solver so that it scales to ten sites and transfers across graphs.
A variational quantum Kolmogorov--Arnold network (VQKAN) generates the angles of a shallow circuit from the time and from features of the distance matrix; the circuit is trained on the conditional value at risk (CVaR) of the tour length with an exact adjoint gradient, a risk-level schedule, multi-start, small-angle initialization, and basin-hopping restarts, and the qubits are read as a Gray code so that neighbouring states are neighbouring tours.
The method returns the optimal tour on every instance tested for $N = 4$ to $10$ (symmetric polygons and asymmetric random graphs), and in 112 of 120 cases of a thirty-run repetition at $N = 10$ with the remaining gaps at 1\% of the optimum, using 3 to 19 qubits where one-hot encoding needs 9 to 81, with as few as $10^2$ measurement shots.
With the distance features as inputs, one parameter set generalizes to unseen times on time-dependent graphs.
An ablation identifies basin hopping as the ingredient that removes the residual local minima at $N = 10$, while deeper circuits do not help; controls against 2-opt, the one-hot encoding, and an $N$-qubit one-hot walk separate the circuit's contribution from classical post-processing.
\end{abstract}

\noindent{\it Keywords\/}: quantum computing, combinatorial optimization, traveling salesman problem, Kolmogorov--Arnold network, variational quantum algorithm

\submitto{Quantum Science and Technology}
\maketitle

\section{Introduction}\label{sec:intro}

The traveling salesman problem (TSP) asks for the shortest closed route through $N$ sites and is NP-hard \cite{lawler1985, cormen2009}.
It underlies routing, scheduling, and circuit layout \cite{appelgate2006, gutin2002}, and exact methods such as branch-and-bound \cite{land1960}, cutting planes \cite{dantzig1954}, and dynamic programming \cite{bellman1962} scale exponentially, so practical solvers rely on local search and metaheuristics \cite{kirkpatrick1983, glover1989, goldberg1989, dorigo1997, Blazinskas2011}.
TSP is also a standard target for quantum optimization: quantum annealing \cite{kadowaki1998, johnson2011, 10247202, dixit_quantum_2023, Smith-Miles_2025}, the quantum approximate optimization algorithm \cite{farhi2014, wang2020, herrman2022}, and variational quantum algorithms (VQAs) on noisy intermediate-scale devices \cite{preskill2018, cerezo2021} have all been applied to it.

The obstacle common to these approaches is the encoding.
The standard Ising formulation assigns one qubit to each (site, position) pair, so a tour needs $N^2$ qubits, or $(N-1)^2$ when the start is fixed \cite{lucas2014ising}.
Most of the $2^{(N-1)^2}$ basis states are not tours, and the Hamiltonian must penalize them; the penalty weights are hard to tune and dominate the landscape.
Binary encodings of the position of each site reduce the count to $(N-1)\lceil \log_2 N \rceil$ qubits but still leave invalid strings that must be penalized or repaired \cite{tabi2020quantum, glos2022space}.
Permutation encodings go one step further: the tour is labeled by its rank among the $(N-1)!$ permutations, for instance by the Lehmer code \cite{lehmer1960teaching}, and stored in $\lceil \log_2 (N-1)! \rceil$ qubits, so that every basis state is a tour and no penalty term exists.
Such encodings have been used with Grover adaptive search \cite{zhu2022realizable}, with quantum dynamic programming \cite{bai2025quantum}, and with variational solvers: Schnaus \emph{et al.} compared them with one-hot and binary encodings under QAOA and VQE and found that the permutation encoding gives the best solutions on small instances because it has no feasibility problem \cite{schnaus2024efficient}, and Fagiolo and Vescera trained a circuit topology on one instance and reused it on others, reaching the optimum in 100\% of samples at four sites but 20\% at seven \cite{fagiolo2025freeze}.
A related encoding stores the tour in the correlations of two $\lceil \log_2 N \rceil$-qubit registers \cite{baldazzi2025resource}.
The open problem with permutation encodings is therefore not the qubit count but the optimization: the landscape over the rank has no locality, every variational run has to be repeated per instance, and the success rate collapses beyond six sites.
Encodings that use only $N$ qubits are possible if the salesman's position is stored as a one-hot vector and moved step by step, but such a state lives in the $N$-dimensional Hamming-weight-one subspace, is classically simulable in $O(N^2)$ operations per step, and cannot represent correlations between successive positions; we quantify this failure in Sec.~\ref{sec:discussion}.
Information-theoretically, $N$ qubits cannot even label all tours once $N \ge 6$, since the read-out of $N$ qubits distinguishes at most $2^N$ outcomes and $(N-1)! > 2^N$ from $N = 6$ on (from $N = 4$ on if the start site is not fixed, since $N! > 2^N$).

Here we address the optimization.
We keep the Lehmer-code encoding, $n = \lceil \log_2 (N-1)! \rceil$ qubits (19 instead of 81 for $N = 10$), and read the qubits as a Gray code so that states one bit flip apart are consecutive ranks, which differ only in the order of the last visited sites.
The circuit is a shallow hardware-efficient ansatz whose rotation angles are produced by a variational quantum Kolmogorov--Arnold network (VQKAN) \cite{2024arXiv240419756L, Wakaura_VQKAN_2024, 2025arXiv250321336W}: each angle is a learnable spline of the input time and of features of the distance matrix, so one parameter set covers a family of graphs whose distances change with time, as in power-grid or traffic routing, without retraining per instance.
The circuit is trained on the conditional value at risk (CVaR) of the tour length \cite{barkoutsos2020cvar}, whose minimizer concentrates probability on the shortest tours and matches a best-of-shots read-out, with an exact adjoint gradient, a schedule of the risk level, multi-start, small-angle initialization, and basin-hopping restarts.

We show that this pipeline returns the optimal tour on every instance we tested for $N = 4$ to $10$, on symmetric polygons and on asymmetric random graphs, and that the read-out needs no more than $10^2$ shots.
We identify which parts of the optimizer are responsible, show that the distance features are what allows generalization to unseen times, and separate the contribution of the circuit from that of classical post-processing with 2-opt, one-hot, and $N$-qubit-walk controls.
Section~\ref{sec:method} describes the encoding, circuit, loss, and optimizer; Sec.~\ref{sec:results} reports the numerical results; Sec.~\ref{sec:discussion} discusses the classical controls and the limitations; Sec.~\ref{sec:conclusion} concludes.

\section{Method}\label{sec:method}

\subsection{Lehmer-code encoding}

Fix site $0$ as the start.
A tour is a permutation $\pi$ of the remaining $N-1$ sites, and its Lehmer code is
\begin{equation}
k(\pi) = \sum_{i=1}^{N-1} d_i \,(N-1-i)!, \qquad 0 \le d_i \le N-1-i,
\end{equation}
where $d_i$ is the number of not-yet-visited sites with a smaller index than the $i$-th visited site.
The map $\pi \mapsto k$ is a bijection onto $\{0, \dots, (N-1)!-1\}$, and $k$ is decoded back to $\pi$ in $O(N^2)$ classical operations by repeatedly dividing by factorials and removing the selected site from the list of unvisited sites.
We store $k$ in $n = \lceil \log_2 (N-1)! \rceil$ qubits; the $2^n - (N-1)!$ integers above $(N-1)!-1$ are reduced modulo $(N-1)!$, so every basis state $|x\rangle$ corresponds to a tour $\pi(x)$ with length
\begin{equation}
L(x) = \sum_{j=0}^{N-1} D_{\pi_j \pi_{j+1}}, \qquad \pi_0 = \pi_N = 0,
\end{equation}
for a distance matrix $D$ that may be asymmetric.
Table~\ref{tab:qubits} compares $n$ with the one-hot and binary-position encodings.
To make neighbouring basis states correspond to neighbouring tours, the integer read from the qubits is interpreted as a binary-reflected Gray code \cite{gray1953pulse} before decoding, so that states one bit flip apart map to consecutive Lehmer codes, which differ only in the order of the last visited sites.

\begin{table}
\caption{Qubit count of TSP encodings with the start site fixed.}\label{tab:qubits}
\begin{tabular}{@{}lccccccc@{}}
\br
$N$ & 4 & 6 & 8 & 10 & 12 & 15 & 20 \\
\mr
one-hot, $(N-1)^2$ \cite{lucas2014ising} & 9 & 25 & 49 & 81 & 121 & 196 & 361 \\
binary position, $(N-1)\lceil\log_2 N\rceil$ \cite{glos2022space} & 6 & 15 & 21 & 36 & 44 & 56 & 95 \\
Lehmer code, $\lceil\log_2 (N-1)!\rceil$ (this work) & 3 & 7 & 13 & 19 & 26 & 37 & 57 \\
\br
\end{tabular}
\end{table}

\subsection{Circuit and VQKAN angle generation}

The state is prepared by $d+1$ layers of single-qubit $R_y$ rotations separated by $d$ layers of nearest-neighbour $CZ$ gates on a line,
\begin{equation}
|\psi(\bm\theta)\rangle = \prod_{l=d}^{0} \Big[ \prod_{q=0}^{n-1} R_y^{(q)}(\theta_{l,q}) \Big] \Big[ \prod_{q} CZ_{q,q+1} \Big]^{[l<d]} |0\rangle^{\otimes n},
\end{equation}
with $d = 2$ throughout.
Because $R_y$ and $CZ$ are real, the amplitudes are real and the probabilities $p_x = \langle x | \psi \rangle^2$ are exactly simulable together with their derivatives by the adjoint method \cite{jones2020efficient}, at the cost of one forward and one backward pass per gradient.

The angles are not free parameters.
Following the VQKAN construction \cite{Wakaura_VQKAN_2024}, each angle is a sum of learnable one-dimensional functions of the input features $\bm z$,
\begin{equation}
\theta_{l,q}(\bm z) = \sum_{f} \phi_{l,q,f}(z_f),
\label{eq:kan}
\end{equation}
where each $\phi_{l,q,f}$ is a quadratic spline with $G = 8$ knots on $[0,1]$, so the trainable parameters are the $n (d+1) \, n_z \, G$ knot values.
The features are the time $t \in [0,1]$ and, when the graph is given to the network, the $N$ row means of $D$ divided by its largest entry; this is a data re-uploading circuit \cite{perezsalinas2020data} whose uploaded values are transformed by splines rather than by fixed linear maps.
Since the spline is linear in its knot values, the Jacobian $\partial \bm\theta / \partial(\text{knots})$ is a constant matrix and the parameter gradient is obtained from the angle gradient by one matrix product.

\subsection{Loss and read-out}

For each input $t$ the loss is the conditional value at risk of the tour length at level $\alpha$ \cite{barkoutsos2020cvar},
\begin{equation}
\mathrm{CVaR}_\alpha(t) = \frac{1}{\alpha} \sum_{x \in \mathcal{T}_\alpha} w_x \, L_t(x), \qquad \sum_{x \in \mathcal{T}_\alpha} w_x = \alpha,
\end{equation}
where $\mathcal{T}_\alpha$ collects the shortest tours until their probability mass reaches $\alpha$ (the boundary state enters with the remaining mass $w_x < p_x$), and $L_t$ is the tour length on the graph $D(t)$.
The total loss is the mean of $\mathrm{CVaR}_\alpha(t)$ over the training inputs.
Its gradient with respect to $p_x$ is $(L_t(x) - L_t(x_b))/\alpha$ for the states fully inside the tail and zero elsewhere, with $x_b$ the boundary state, which is exact on the current tail and is what the adjoint pass propagates.

The read-out mimics the measurement on hardware: the state is sampled $M$ times, each sample is decoded to a tour, and the shortest sampled tour is returned.
In the state-vector simulation this is the shortest tour among the states with $p_x \ge 1/M$; we use $M = 10^3$ in the main tables and vary $M$ in Sec.~\ref{sec:shots}.
As an optional classical post-processing step we apply 2-opt \cite{croes1958method} to each of the 20 shortest readable tours and keep the best; because $D$ may be asymmetric, each 2-opt move is evaluated by recomputing the full tour length.

\subsection{Optimizer}\label{sec:optimizer}

The parameters are trained by L-BFGS-B \cite{liu1989limited} with the exact gradient.
Four measures address the local minima of the CVaR landscape.
(i) Multi-start: $K$ independent runs from different random seeds, keeping the run with the lowest final loss.
(ii) Risk-level schedule: the run proceeds through $\alpha = 0.5, 0.2, 0.1, 0.05, 0.02$, warm-starting each stage from the previous one, so the landscape is smooth early and concentrated on the optimum late.
(iii) Small-angle initialization: knot values are drawn from $U(-0.1, 0.1)$, so the circuit starts near the identity and away from the flat region of a random deep circuit \cite{mcclean_barren_2018}.
(iv) Basin hopping \cite{wales1997global}: after each run converges, the parameters are perturbed by $\mathcal{N}(0, 0.3^2)$, re-optimized at the two smallest $\alpha$, and accepted if the loss decreases; this is repeated $H$ times.
We use $K = 8$, $H = 3$, and 60 L-BFGS-B iterations per stage for $N \le 8$, and $K = 3$, $H = 2$, 40 iterations for $N = 10$.
The single-start reference configuration in the tables uses one seed, fixed $\alpha = 0.2$, knot values from $U(0, \pi)$, no hopping, and the time $t$ as the only feature.

\subsection{Instances and protocol}

For each $N \in \{4, 6, 8, 10\}$ we use three graphs: a regular polygon with unit-length long chords and side length $1/\sqrt{2}$ ($N = 4$, diagonal 1) or $0.5$ ($N \ge 6$, all other distances 1), and two asymmetric random graphs with $D_{ij} \sim U(0.05, 1.05)$.
Each graph is solved in three independent runs of the optimizer, with the loss averaged over four network inputs $t \in \{0, 0.1, 0.2, 0.3\}$; on these static graphs the four inputs see the same $D$, so a run counts as successful at an input if the read-out equals the optimum found by exhaustive enumeration of the $(N-1)!$ tours.
We report the number of successful (graph, run, input) triples out of 36 and the mean gap between the returned and optimal lengths.
All simulations are exact state-vector simulations written in NumPy, with the circuit also constructed in blueqat \cite{blueqat} for cross-checking.

\section{Results}\label{sec:results}

\subsection{Static graphs, $N = 4$ to $10$}

\begin{table}
\scriptsize
\caption{Static graphs. Fraction of (graph, run, input) triples that returned the optimal tour, out of 36, and mean length gap to the optimum. ``Full optimizer'' is the method of Sec.~\ref{sec:optimizer} with the distance features; ``single start'' is the reference configuration. ``2-opt from random'' starts 2-opt from a random tour. The $N$-qubit one-hot walk is described in Sec.~\ref{sec:discussion}; at $N \ge 8$ it was not run.}\label{tab:main}
\begin{tabular}{@{}ccccccc@{}}
\br
$N$ & qubits & parameters & full optimizer & single start & 2-opt from random & $N$-qubit one-hot walk \\
 & Lehmer / one-hot & full / single & optimal (gap) & optimal (gap) & optimal (gap) & optimal (valid) \\
\mr
4  & 3 / 9   & 360 / 72   & 36/36 (0.000) & 36/36 (0.000) & 29/36 (0.105) & 12/36 (36/36) \\
6  & 7 / 25  & 1176 / 168 & 36/36 (0.000) & 36/36 (0.000) & 15/36 (0.430) & 0/36 (0/36) \\
8  & 13 / 49 & 2808 / 312 & 36/36 (0.000) & 11/36 (0.245) & 15/36 (0.320) & -- \\
10 & 19 / 81 & 5016 / 456 & 36/36 (0.000) & 7/36 (0.434)  & 12/36 (0.633) & -- \\
\br
\end{tabular}
\end{table}

\begin{figure}[t]
\includegraphics[width=\textwidth]{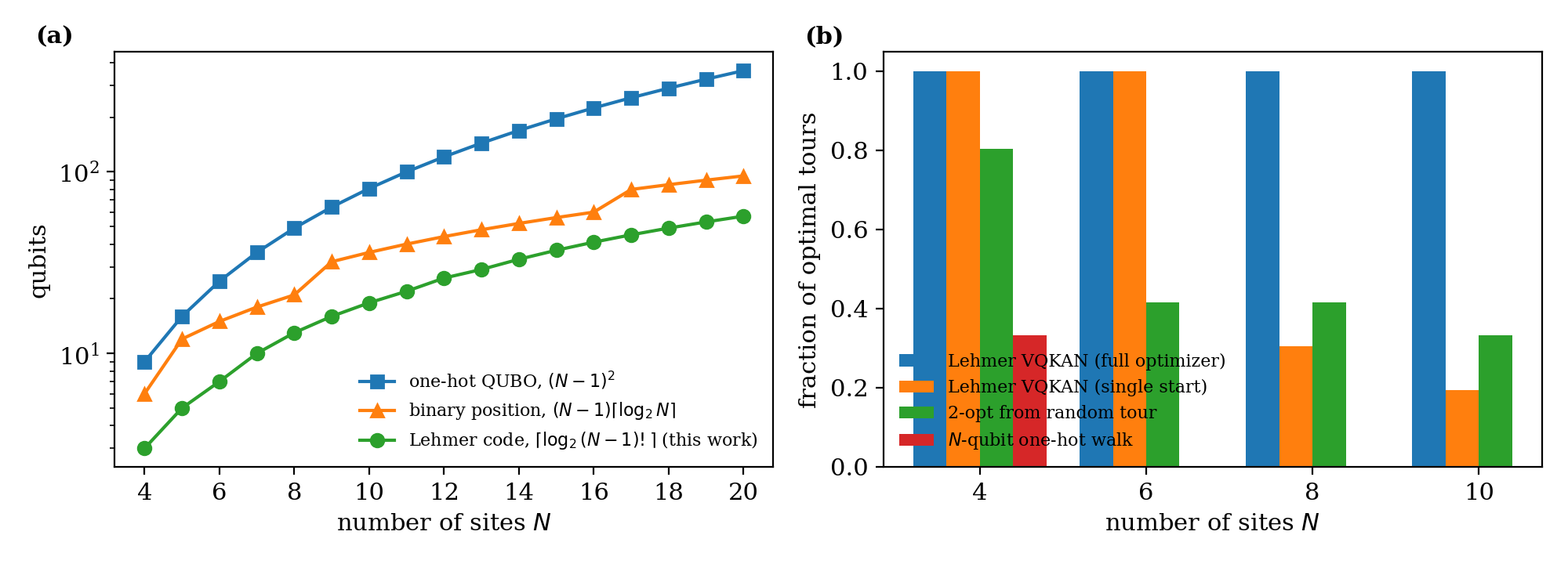}
\caption{(a) Qubit count of the three encodings of Table~\ref{tab:qubits}. (b) Fraction of optimal tours on the static graphs of Table~\ref{tab:main}.}\label{fig:summary}
\end{figure}

Table~\ref{tab:main} and Fig.~\ref{fig:summary} summarize the static-graph results.
The full optimizer returns the optimal tour in all 36 cases at every $N$.
The single-start configuration is sufficient up to $N = 6$ but falls to 11/36 at $N = 8$ and 7/36 at $N = 10$; its failures are local minima of the CVaR loss in which a tour with one or two long edges carries the tail mass.
2-opt started from a random tour, the classical control with the same post-processing budget, reaches the optimum in 29/36 cases at $N = 4$ and in 12 to 15 out of 36 at $N \ge 6$, so the trained circuit is not reproducing what 2-opt alone would find.
When 2-opt is instead started from the read-out of the single-start circuit, the mean gap drops from 0.245 to 0.154 at $N = 8$ and from 0.434 to 0.151 at $N = 10$, compared with 0.320 and 0.633 from random starts: even the under-optimized circuit provides a better starting tour than a random one.
Over the thirty-run repetition at $N = 10$ reported in Sec.~\ref{sec:ablation}, the full optimizer returned the optimum in \xtenBfiveOpt{} of 120 cases and in 28 of the 30 runs at every input.

\subsection{Which parts of the optimizer matter}\label{sec:ablation}

\begin{table}
\caption{Ablation at $N = 10$ (36 triples each). ``A'' is the single-start configuration plus multi-start, the $\alpha$ schedule, small-angle initialization, and the 20-state 2-opt read-out. ``B'' adds a depth-4 circuit with alternating nearest-neighbour and range-$n/2$ $CZ$ layers, the Gray-coded index, the distance features, and basin hopping. Each ``B$-$'' row removes one of these four. The row ``B$-$depth'' is the configuration used in Table~\ref{tab:main}.}\label{tab:ablation}
\begin{tabular}{@{}lccc@{}}
\br
configuration & parameters & optimal (gap) & with 2-opt \\
\mr
single start & 456 & 7/36 (0.434) & 14/36 (0.151) \\
A & 456 & 16/36 (0.044) & 22/36 (0.012) \\
B & 8360 & 32/36 (0.002) & 32/36 (0.002) \\
B $-$ depth-4 circuit & 5016 & 36/36 (0.000) & 36/36 (0.000) \\
B $-$ Gray code & 8360 & 28/36 (0.004) & 28/36 (0.004) \\
B $-$ distance features & 760 & 28/36 (0.006) & 30/36 (0.005) \\
B $-$ basin hopping & 8360 & 20/36 (0.048) & 20/36 (0.032) \\
\br
\end{tabular}
\end{table}

Table~\ref{tab:ablation} isolates the contribution of each measure at $N = 10$, where the single-start configuration fails most often.
Multi-start, the $\alpha$ schedule, and the small-angle initialization together (row A) reduce the mean gap by an order of magnitude at fixed parameter count, but the remaining failures are a single local minimum that all seeds share on each random graph, at 1 to 3\% above the optimum.
Basin hopping is what removes it: taking it away from B costs 12 of 36 cases and returns the gap to the level of A, whereas removing the Gray code or the distance features costs 4 cases each with gaps below 0.01.
The deeper circuit does not help: removing it solves all 36 cases with 40\% fewer parameters.
At $N = 8$ every one of the seven configurations except the single start and A returns 36/36, so the ranking is visible only at $N = 10$.
To resolve the two leading configurations we repeated them with ten runs per graph (30 runs, 120 cases each): the depth-2 configuration returned the optimum in \xtenBfiveOpt{} cases (mean gap \xtenBfiveGap, largest gap 0.016) and the depth-4 configuration in \xtenBOpt{} (mean gap \xtenBGap, largest gap 0.016), all misses being on the same random graph and 1\% above the optimum.
The four-case difference is one run and does not separate the two; the depth-2 circuit is preferred for its parameter count.

\subsection{Time-dependent graphs}\label{sec:timedep}

\begin{table}
\caption{Time-dependent graphs $D(t) = (1-t) R_0 + t R_1$ with $R_0, R_1$ asymmetric random. Three families per $N$; the optimal tour changes 2--3 times ($N = 6$) or 4--6 times ($N = 8$) along $t \in [0, 1]$. Trained on $t \in \{0, 0.25, 0.5, 0.75, 1\}$ (15 cases), tested on $t \in \{0.125, 0.375, 0.625, 0.875\}$ (12 cases).}\label{tab:timedep}
\begin{tabular}{@{}clccc@{}}
\br
$N$ & features & train & test (gap) & test with 2-opt \\
\mr
6 & $t$ and distances & 15/15 & 11/12 (0.013) & 12/12 \\
6 & $t$ only           & 15/15 & 4/12 (0.203)  & 8/12 \\
8 & $t$ and distances & 15/15 & 4/12 (0.133)  & 11/12 \\
8 & $t$ only           & 15/15 & 0/12 (0.372)  & 3/12 \\
\br
\end{tabular}
\end{table}

The purpose of generating the angles by a network rather than optimizing them directly is to cover a family of graphs with one parameter set.
Table~\ref{tab:timedep} tests this on families whose optimal tour changes several times as $t$ runs from 0 to 1.
Both feature sets fit the five training times exactly.
At the four held-out times the network that only receives $t$ interpolates poorly, 4/12 at $N = 6$ and 0/12 at $N = 8$, while the network that also receives the row means of $D(t)$ reaches 11/12 and 4/12, and 12/12 and 11/12 after 2-opt.
The distance features, whose effect on static graphs in Table~\ref{tab:ablation} was small, are therefore what carries the generalization across time.

\subsection{Finite measurement shots}\label{sec:shots}

The tables above read out the state with $M = 10^3$ shots.
Sampling the trained states of Table~\ref{tab:main} with $M = 10^2$, $10^3$, $10^4$, and $10^5$ shots (three sampling seeds per state) returned the optimal tour in 100\% of cases at every $M$ for $N = 4$, 6, and 8.
For the 30 trained states of the $N = 10$ repetition the rate is 93.1\% at $M = 10^2$ and 93.3\% at $M \ge 10^3$, equal to the exact-read-out rate of 112/120: the two runs that miss the optimum do so because the state does not contain it, not because of sampling.
This follows from the final risk level: at $\alpha = 0.02$ the optimizer places at least 2\% of the probability on the shortest tours, and in practice much more, so a few hundred shots contain the optimum with high probability.
The training itself uses exact probabilities; the effect of shot noise on the gradient is not studied here.

\subsection{One-hot control at $N = 4$}

To check that the gains come from the encoding rather than from the optimizer, we ran the same optimizer, circuit, and read-out on the one-hot encoding with $(N-1)^2 = 9$ qubits at $N = 4$, assigning a penalty length of twice the longest tour to invalid basis states.
It also returned the optimal tour in 36/36 cases, but with 1080 parameters instead of 360, and with 55\% of the read-out probability on invalid strings.
At $N = 6$ the one-hot state space has $2^{25}$ dimensions against $2^7$ for the Lehmer code, and only $120$ of the $3.4 \times 10^7$ basis states are tours; we did not simulate it.

\section{Discussion}\label{sec:discussion}

\subsection{Why $N$ qubits are not enough}
An encoding that uses one qubit per site and stores the salesman's current position as a one-hot vector, moving it with number-conserving two-qubit rotations at each of $N$ steps, uses the smallest qubit count of all, $N$.
We implemented this walk with the same VQKAN angle generation, reading the position at each step as the expectation values $\langle Z_j \rangle$ and building the loss from products of expectation values at successive steps plus a term that penalizes returning to a visited site.
Its results are in the last column of Table~\ref{tab:main}: at $N = 4$ it returned the optimum only on the symmetric square, and the same tour on all three graphs; at $N = 6$ it returned no valid tour.
The reason is structural.
The state never leaves the Hamming-weight-one subspace, which has dimension $N$, so the entire computation is a sequence of $N$-dimensional real rotations and is classically simulable in $O(N^2)$ operations per step; a loss built from single-step marginals $\langle Z_j \rangle_t \langle Z_k \rangle_{t+1}$ is a mean-field relaxation that cannot represent the joint distribution of successive positions.
The Lehmer encoding uses $\log_2 N!$ rather than $N$ qubits precisely so that the state space is large enough to hold a distribution over tours.

\subsection{Relation to other permutation-encoded solvers}
The closest work is that of Fagiolo and Vescera \cite{fagiolo2025freeze}, who also use a Lehmer-type encoding and also aim at reuse across instances: the circuit topology is found by simulated annealing on one instance, frozen, and only the parameters are re-optimized on a new instance, giving optimal-tour sampling rates of 100\%, 90\%, 80\%, and about 20\% at 4, 5, 6, and 7 sites on symmetric random graphs.
Our pipeline differs in three respects.
The topology is fixed (a depth-2 ladder) and what is learned is a map from graph features to angles, so a new time point needs no re-optimization at all (Table~\ref{tab:timedep}); the optimizer uses exact gradients and explicit escape moves rather than annealing over topologies; and the rank is read as a Gray code.
With these, the success rate at $N = 8$ and $N = 10$ is 100\% and 93\% of cases, on asymmetric as well as symmetric graphs, where the frozen-ansatz approach drops to 20\% at $N = 7$.
Schnaus \emph{et al.} \cite{schnaus2024efficient} report that the permutation encoding has the best solution quality among the encodings they compare under QAOA and VQE on small instances; our results extend this observation to $N = 10$ and show that its remaining weakness, the local minima of a landscape without locality, can be handled on the optimizer side.

\subsection{What the simulation does and does not show}
The state-vector simulation enumerates all $(N-1)!$ tour lengths to evaluate the loss, so it establishes the qubit count, the absence of penalty terms, and the behaviour of the optimizer, but not a quantum advantage.
On hardware the loss would be estimated from samples; Sec.~\ref{sec:shots} shows that the read-out is robust to $10^2$ shots, while the gradient estimation under shot noise remains to be tested.
The 2-opt control shows that the circuit does more than classical local search at $N \le 6$ and provides better starting tours than random ones at $N \ge 8$; a fair comparison with state-of-the-art classical heuristics at larger $N$ is outside the scope of this work.

\subsection{Statistics}
Each configuration in Table~\ref{tab:ablation} rests on three runs per graph, so differences of four cases, such as those between the Gray-code and distance-feature ablations, are within one run and are not resolved.
The effect of basin hopping (12 to 16 cases) and the ranking of the single-start configuration are robust.
At $N = 10$ the thirty-run repetition in Sec.~\ref{sec:ablation} puts the success rate of the proposed configuration at 93\% of cases (112/120) with all misses within 1\% of the optimum, and shows that the depth-4 circuit is not distinguishable from it.

\subsection{Scaling}
The qubit count $\lceil \log_2 (N-1)! \rceil \approx N \log_2 N - 1.44 N$ grows as $N \log N$; for $N = 20$ it is 57, within reach of current devices, where the one-hot encoding needs 361.
The classical decoding of a sample costs $O(N^2)$ and the 2-opt post-processing $O(N^2)$ per pass, both negligible.
The optimizer cost is dominated by the number of circuit evaluations, $K (H+1)$ runs times five $\alpha$ stages; at $N = 10$ a single run takes minutes on a laptop in simulation.

\section{Concluding Remarks}\label{sec:conclusion}

We have taken the Lehmer-code encoding of the traveling salesman problem, $\lceil \log_2 (N-1)! \rceil$ qubits with every basis state a tour and no constraint term, and trained a shallow circuit on it whose angles are generated by a variational quantum Kolmogorov--Arnold network from the time and the distance matrix, on the CVaR of the tour length with an exact adjoint gradient.
With multi-start, a risk-level schedule, small-angle initialization, and basin hopping, the method returned the optimal tour on all symmetric and asymmetric instances from $N = 4$ to $N = 10$ in the three-run protocol, and in 112 of 120 cases with the rest within 1\% in a thirty-run repetition at $N = 10$, using 3 to 19 qubits and read-out from as few as $10^2$ shots.
The network's distance features are what lets one parameter set follow a graph whose distances change with time.
The next steps are training under shot noise, larger $N$ where the classical enumeration of tours is no longer available and the loss must be sampled, and a comparison with classical heuristics beyond 2-opt.

\section*{Data availability}
The simulation code, the result files behind every table, and the scripts that regenerate them are available at \url{https://github.com/WakauraH/tsp-lehmer-vqkan}.

\section*{Conflict of interest}
The author has no conflicts to disclose.

\section*{References}
\bibliographystyle{iopart-num}
\bibliography{mainwakaura2,extra_v2}
\end{document}